\documentclass[%
reprint,
 amsmath,amssymb,
 aps,
]{revtex4-2}

\usepackage{graphicx}
\usepackage{dcolumn}
\usepackage{bm}

\usepackage{ulem}
\usepackage{xcolor}
\usepackage{subfig}

\usepackage{verbatim} 

\newcommand{\beginsupplement}{%
        \setcounter{table}{0}
        \renewcommand{\thetable}{S\arabic{table}}%
        \setcounter{figure}{0}
        \renewcommand{\thefigure}{S\arabic{figure}}%
     }

\begin{document}

\preprint{APS/123-QED}

\title{On-lattice Vicsek model in confined geometries}

\author{Andreas Kuhn* }
\affiliation{andreas.kuhn@uni-wuerzburg.de}

\author{Sabine C. Fischer}%
\affiliation{sabine.fischer@uni-wuerzburg.de \\
\\ Center for Computationaland Theoretical Biology, Fakultät für Biologie, Universität Würzburg, Klara-Oppenheimer-Weg 32, 97074 Würzburg, Germany }%

\date{\today}

\begin{abstract}
The Vicsek model (Vicsek et al. 1995) is a very popular minimalist model to study active matter with a number of applications to biological systems at different length scales. With its off-lattice implementation and periodic boundary conditions, it aims at the analysis of bulk behaviour of a limited number of particles. We introduce an efficient on-lattice implementation with finite particle volume and analyse its behaviour for three different geometries with reflective boundary conditions. For sufficiently fine lattices, the model behaviour does not differ between off-lattice and on-lattice implementation. The reflective boundary conditions introduce an alignment of the particles with the boundary for low levels of noise. Numerical sensitivity analysis of the swarming behaviour results in a detailed characterisation of the on-lattice Vicsek model for confined geometries with reflective boundary conditions. In a channel geometry, the boundary alignment causes swarms to move along the channel. In a box, the edges act as swarm traps and the trapping shows a discontinuous noise dependence. In a disk geometry, an ordered rotational state arises. This state is well described by a novel order parameter. Our works provides a foundation for future studies of Vicsek-like models with discretized space.

\end{abstract}

\maketitle


\section{\label{sec:level1}Introduction}

Swarms of fish, flocks of birds or herds of big mammals are spectacular macroscopic phenomena that are familiar to everybody. At first glance, they appear to be the result of a complex interplay between highly developed animals. But past research has shown that such behaviour can be described by simple interaction rules that each individual follows \cite{ vicsek_nature}. Biological systems with similar properties can be found at much smaller length scales including microorganisms \cite{Bacteria_collective_motion_1,Bacteria_collective_motion} and even subcelullar components \cite{Sciortinoe2017047118}. However, collective behaviour is not limited to living organisms. Micromotors as well as macroscopic and granular rods also exhibit collective behaviour  \cite{Light_powerd_nanamotor,collective_metallic_rods,collective_metallic_rods_2}. Surprisingly, systems with collective behaviour show effects that have previously only been known for thermodynamic equilibrium systems \cite{ vicsek_nature}. For example, migrating tissue cells can undergo a phase transition from a disordered to an ordered state \cite{phase_transition_tissue_cells}.

A complete picture of many of these so called active matter systems does not exist yet, but the results of ongoing research underline the universal properties and the close connection of collective behaviour to statistical physics. The Vicsek model (VM) \cite{Vicsek} is one of the most prominent theoretical models to study active matter. It is able to create complex macroscopic behaviour like swarming or phase transitions \cite{Vicsek_final_understood} through a very simple velocity alignment rule for the interaction of particles with their neighbours. It has been applied to a wide range of systems and different variants including alterations to the interaction rule and an extension to a continuous model have been implemented \cite{CZIROK200017}. Numerical simulations for the basic VM are implemented off-lattice on a square simulation domain with periodic boundary conditions. Such a setup is useful for approximating bulk properties from finite size simulations. However, collective behaviour in many biological systems relies on a large number of individuals interacting in a confined environment. In this case, the effects of the geometry and the boundaries of the environment on the behaviour of a large number of particles have to be examined by direct simulations.

Including local repulsive forces in the basic VM limits the maximal local density in the system. In this adaptation of the VM, confining the particles by a circular reflective boundary results in rotations of the particles both clockwise and anti-clockwise in the high density and low noise regime \cite{CZIROK200017}. A similar behaviour has been observed for the continuous version without local repulsion of the VM \cite{ARMBRUSTER201758}. Extending the analysis to a channel with two parallel periodic and two reflective boundaries and a box with four reflective boundaries reveals shape independent features of the collective behaviour. For all three geometries, the boundaries introduce a spatial coherence of swarms in the continuous VM that has not been observed in the basic version. Hence, in finite domains of reasonable size, the boundaries may act as attractors. 

Analysis of the collective behaviour of repulsive discrete particles in the same three shapes shows a different picture \cite{Hiraoka_2017}. Spatially incoherent transition states are observed, where particles are ordered and densely packed in one part of the confined environment and the remaining system shows disorder. After a slow transition, phase diagrams for order parameters adapted to the different geometries show that these states are eventually resolved into coherent behaviour as for periodic boundary conditions. This leads to the hypothesis that the boundaries hinder the spreading of the correlations.

These observations point to the importance of a more narrow study of the effects of boundary conditions together with a local repulsion due to a finite particle volume. Therefore, we employ an on-lattice implementation with a finite particle volume. As geometries, we consider a channel, a box and a disk with reflective boundary conditions and analyse their influence on the collective behaviour of the particles. We find that for a sufficiently small lattice, the behaviour of the VM is independent of the type of implementation. The reflective boundary conditions result in particle alignment parallel to the tangential vector of the closest boundary point for sufficiently low noise levels in all geometries. For high levels of noise, the particle movement is uncorrelated similar to the basic VM. The boundary alignment results in behaviour particular to each geometry. For the channel, separate swarms moving in opposite directions at the top and the bottom wall can occur. In the box geometry, the boundary alignment can yield particle trapping in the corners. For the disk geometry, the boundary effect results in an ordered rotational state. Characterisation of the VM for these different geometries extends its applicability to a wider range of biological systems. 

\section{Model and order parameter}
\label{Model}
A system in the basic VM \cite{Vicsek} consists of $N$ particles with two properties: position $\vec{x}_i(t)$ and velocity $\vec{v}_i(t)$. All particles move with the same absolute velocity $v_0$. The direction of movement is expressed by the angle $\Theta_i(t)$. Hence, the velocity is described by
\begin{equation}
\vec{v}_i(t) =v_0 \begin{pmatrix}\cos\Theta_i(t) \\\sin\Theta_i(t)
\end{pmatrix}
\end{equation}
   
\begin{figure}[htbp]
\begin{center}
\includegraphics[width=0.8\linewidth]{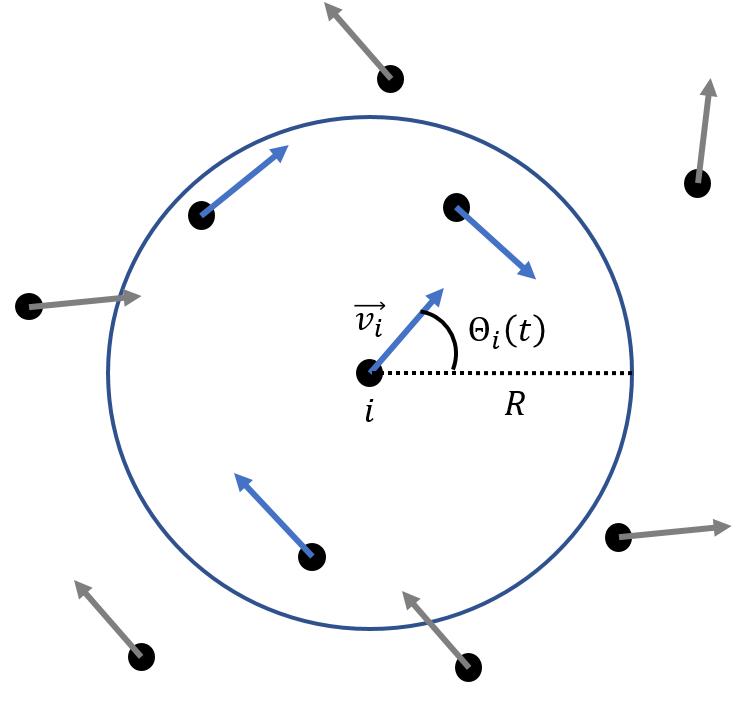}
\caption[]{Alignment mechanism in the VM. The black dots mark the current position of the particles and the arrows point towards their current direction of movement. The i-th particle aligns its direction with all particles inside its interaction radius $R$ (indicated by the blue circle and the blue arrows). }   
\label{fig:alignment}
\end{center}
\end{figure}

All particles exist in a square shaped system (edge length L) with periodic boundary conditions. In the initialisation phase, the system gets populated with $N$ particles with randomly assigned positions and directions of movement. Afterwards, the model evolves in discrete time steps $\Delta t = 1$. During each time step, two updates are conducted for each particle:  

1. Update of the particle direction. Each particle is assigned a new direction 
\begin{equation}
\Theta_i(t+\Delta t) = \langle \Theta_j(t) \rangle_{|x_i-x_j|<R}+\zeta_i(t),
\label{eq:direction}
\end{equation}

with
\begin{equation}
\langle \Theta_j (t) \rangle = \arctan\frac{\langle \sin\Theta_j(t)\rangle}{\langle \cos\Theta_j(t)\rangle},  
\end{equation} 

where  $\langle \Theta_j(t) \rangle_{|x_i-x_j|<R}$ is the average direction of all particles $j$ surrounding the $i$-th particle (including itself) within a radius $R$ (Fig.~\ref{fig:alignment}).  The parameter $R$ is the interaction range of each particle. This alignment term creates order in the system.

The second term $\zeta_i(t)$ is a random angle drawn from a uniform probability distribution over ($-\eta/2,\eta/2$). The range of the distribution $\eta$ can be interpreted as a temperature of the system. It is the main mechanism that counteracts the alignment and hence the order in the system.

2. After the directions are updated for all particles the positions get updated. Each particle is assigned a new position according to
\begin{equation}
x_i(t+\Delta t) = x_i(t)+ v_0\Delta t \begin{pmatrix}\cos\Theta_i(t) \\\sin\Theta_i(t)
\end{pmatrix}.
\label{eq:next_step}
\end{equation}

In summary, the VM has five free parameters, particle number $N$, system size $L$, velocity $v_0$, interaction range $R$ and the noise strength $\eta$. These parameters are not independent of each other and can be reduced to four effective parameters. The behaviour of the VM only depends on the system size $\frac{L}{R}$, the noise strength $\eta$, the density of interaction spheres $\rho = \frac{NR^2}{L^2}$ and the ratio of velocity and interaction range $v_r =\frac{v_0}{R}$.

To quantify the behaviour of the model, Vicsek et al. \cite{Vicsek} have introduced the polar order parameter $v_a$, such that
\begin{equation}
v_a = \frac{1}{Nv_0}\Bigg|\sum_{i=1}^{N}\vec{v}_i\Bigg|.
\label{eq:order_parameter}
\end{equation}

The order parameter $v_a$ is the average normalised velocity of the system. That means, if $v_a = 1$, all particles move in the same direction, and if $v_a \approx  0$ all particles move uncorrelated in the system. It should be added that the order parameter can only truly reach zero in infinite systems, as there are finite size effects in the direction summation which lead to a remaining order parameter $v_a \approx \frac{1}{\sqrt{N}}$ \cite{Vicsek_final_understood} even in  total random systems. 
The VM shows a phase transition, from an ordered motion state to a disordered motion state, with increasing $\eta$ \cite{vicsek_model_vectorial_noise}. In large systems, the state of the system changes from coherently moving swarms, to a band phase to a "cross sea" phase and finally to a completely uncorrelated state \cite{K_rsten_2020}.

\section{On-lattice-hybrid implementation} \label{used imp}
The original and many of the previous implementations of the VM use an off-lattice model \cite{Vicsek,Phase_transition_2009,Chate_2008}. To update the direction of each particle (eq.~\ref{eq:direction}), it is necessary to take all particles into account that are at most a distance $R$ away. In an off-lattice implementation, this requires pairwise comparison of all particles, i.e. the computation time scales with  $N^2$. Hence, pure off-lattice implementations are not suitable to simulate systems with a large number of particles. The typical workaround to this problem is to divide the simulation space in smaller boxes with size $  \geq R$ \cite{Vicsek_final_understood}. Therefore, a pairwise comparison is only required for all particles inside the same and the neighbouring boxes. The computation time scales with $N$ in this improved implementation. 

We chose an alternative workaround to the $N^2$ scaling problem. We implemented an on-lattice implementation where the simulation space is discretized to a two-dimensional lattice and each particle is occupying one lattice site. To update the direction of each particle (eq.~\ref{eq:direction}) in this system, only the grid points within the interaction range of each particle need to be checked. Hence, the computation time scales with $N$. One advantage of this approach is that a finite particle volume (one lattice site) is inherently included in this implementation without adding additional complexity and computation time. Therefore, an additional step is included into the update process. The calculated next position by eq.\ref{eq:next_step} is rounded to the nearest gird point (smallest euclidean distance). If said grid point is already occupied by another particle, then the particle stays at its original position (but keeps its updated direction). Another advantage is the simple expandability, as additional parameters like velocity damping, different boundary conditions or a flow field can be encoded locally in each grid point. Such additional parameters would be naturally read out by the update algorithm in each time step and therefore would only require minor changes to the model. However, one disadvantage is that the precision of an on lattice model is limited by available system memory. To mitigate this problem an ”on-lattice-hybrid” model was created.

\begin{figure} 
\begin{center}
	\includegraphics[width=1.0\linewidth]{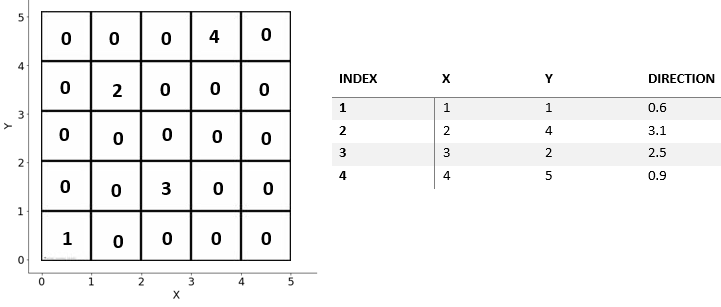}
\caption{ Illustration of the memory layout of the on-lattice-hybrid implementation with a grid size of 5 and a particle number of 4. The entry in the grid cell (left) is the particle ID, where 0 indicates an empty cell. The help array (right) contains the ID, the grid position and the direction of movement for each particle.}
	\label{fig:comparision_lattice}
	\end{center}
\end{figure}

As in a pure on-lattice implementation, the ID and (implicitly) the position of each particle are stored in a discrete two-dimensional grid. All further particle parameter values are saved in a help-array (Fig.~\ref{fig:comparision_lattice}). Due to the necessary homogeneous data structure in the grid, the on-lattice-hybrid implementation requires only half the memory for empty grid cells compared to a pure on-lattice implementation. This yields a significant reduction in required memory, if more grid points are empty than occupied. This condition is easily fulfilled with typical model parameter values. Setting e.g. $v_0 =5$, $R =18$ and $\rho = 1$ results in a grid cell occupation probability of $\approx 0.003$.

To provide a good intuition on how the system is discretized and how precise it is, we chose to deviate from the dimensionless parameter notation mentioned in sec. \ref{Model} and give $v_0$, $R$ and $L$ in units of lattice sites. If the lattice is sufficiently fine ($v_0 \geq 2$ and $R \geq 2$), the simulation results agree with the results for an off-lattice implementation  (see Appendix \ref{Comparison_on_off} and Fig.~\ref{fig:comparision_lattice_2}). Unless mentioned otherwise, all data in the following sections have been determined after a sufficient amount of time to ensure that the corresponding systems have left their initialisation phase, and did not show any significant temporal variations in the studied quantities. 

\section{Reflective boundaries} 
The basic VM uses periodic boundary conditions (PBCs). This is well suited to analyse the bulk behaviour in large systems, but has limited value for smaller sized systems including essentially all lab experiments. In active matter systems, boundaries can have a defining role on particle behaviour \cite{boundary_flow_field}. This can lead to surprising effects, including shape dependent pressure \cite{force_boundary}.

Similar to previous approaches \cite{ARMBRUSTER201758,Hiraoka_2017}, we modeled the boundaries of the simulation space as static walls, and the collisions as elastic. Hence, the incidence angle equals the emergent angle and the absolute value of the velocity $v_0$ does not change upon reflection. Algorithmically this is handled by tracing the path of each particle during position update (eq. \ref{eq:next_step}). If there is a boundary on this path, the particle gets reflected and "walks" the remaining length of the original path in the direction of the newly calculated reflected path. In the following sections, we analyse the influence of the reflective boundary conditions in a channel, a box and a disk.


\subsection{Channel} \label{Channel}
\begin{figure}[ht!]
	\subfloat[\label{channel:low_noise}] {%
		\includegraphics[ width= 0.48\linewidth ]{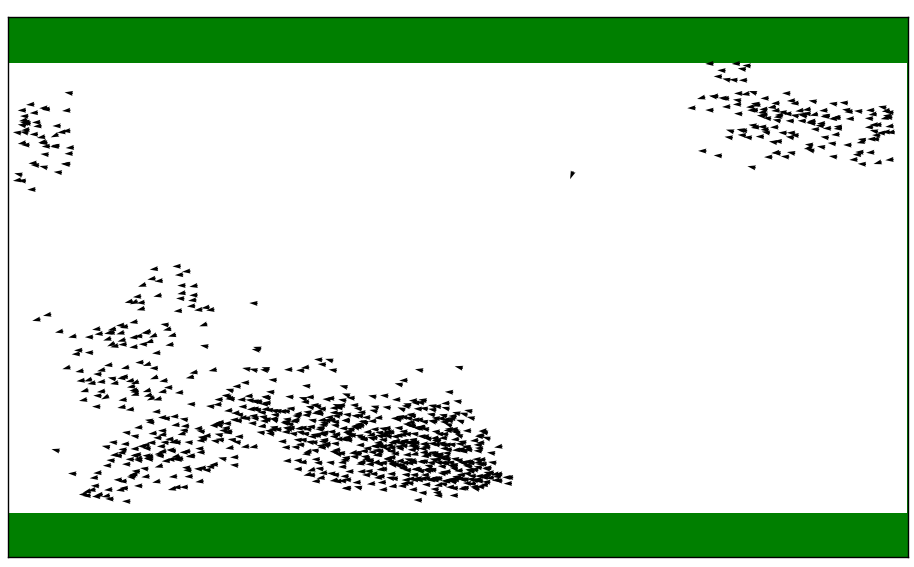}}
	\hspace{\fill}
	\subfloat[\label{channel:high_noise}]{%
		\includegraphics[ width=0.48\linewidth ]{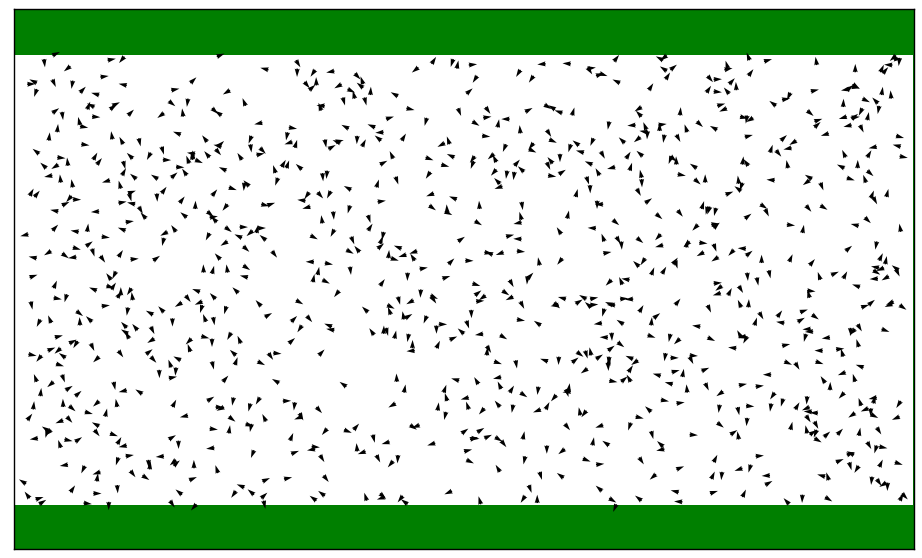}}
	\hspace{\fill}
	\caption{\label{fig:snapshot_channel}Snapshots of simulations of the VM in a channel geometry after 500 time steps with $N = 1000, v_0 = 5, L = 500, R = 18$. (a) For low noise level $\eta$ = 0.2 and (b) for high noise level $\eta$ = 4.2. The reflective boundaries in the y-direction are depicted as green walls. The arrow heads indicate the current movement direction of the individual particles.}
	
\end{figure}

The channel geometry has periodic boundaries in the x-direction and reflective boundaries in the y-direction. Simulations of the VM in the channel for different noise levels show a similar ordering behaviour as for a system with full PBCs. At low noise, swarms are forming and move coherently in one direction and at high noise the particles move uncorrelated (Fig \ref{fig:snapshot_channel}). The main difference is that stable swarms occur only parallel to the boundaries. Hence, in comparison to the VM with full PBCs, the ordered states in the channel geometry become simpler, because long-lived  correlations of orientation can only occur parallel to the walls. The absence of other possible stable states like e.g. a bouncing between the boundaries, can be explained by the following observation on the effect of the Vicsek alignment (eq.~\ref{eq:direction}) on swarm behaviour at the boundary:

When a swarm collides with a reflective boundary, the first incoming particles get reflected, and start to move away from the boundary. In the next time step, the reflected particles align themselves with all particles inside their interaction area (circle with radius $R$). Most of these particles are still moving towards the boundary. Therefore, the movement direction of the outgoing particles is changed quite drastically due to the alignment. The reflected particles move again towards the boundary, but now with a smaller incidence angle. In the same manner, the incoming particles decrease their incidence angle through alignment with the outgoing particles. After some iterations of this process, the whole swarm is moving parallel to the boundary.

Considering a solitary swarm, the sum of incoming and outgoing particles for all alignment processes over all time steps during a collision of the swarm with a boundary is zero. Therefore, the perpendicular velocity component of the swarm towards the boundary, which is opposite for incoming and outgoing particles, is subsequently cancelled, and only the parallel velocity component persists. After some time, all the swarms have collided with a boundary, and consequently all particles move parallel to the boundaries.

\begin{figure}[htbp] 
	\centering
	\includegraphics[width=\linewidth]{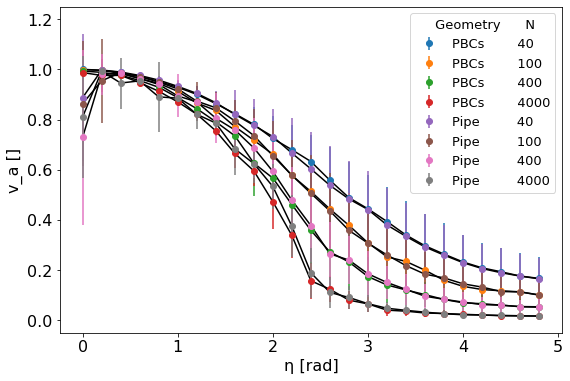}
	\caption{Order parameter versus noise strength $\eta$ for systems with PBCs and a channel geometry, with increasing size and particle number but constant density $\rho = 1$, $v_0 = 5$ and $R =18$. Data points are the mean values with standard deviations of 30 runs each. The black lines connect the data points for a better visual perception of the transitions.}
	\label{fig:comparison_channel_PBCs}
	\end{figure}

The order parameter $v_a$ from the base model (eq.~\ref{eq:order_parameter}) is also suitable to describe the transition from disorder to order in the channel geometry. The behaviour of $v_a$ versus the noise $\eta$, is almost identical for systems with a channel geometry or PBCs (Fig.~\ref{fig:comparison_channel_PBCs}). Only zero noise systems in a channel geometry show deviations from perfect alignment. In this case, the system can be "trapped" in a state, for which the swarms aligned to the top wall and to the bottom wall are completely isolated from each other. These isolated swarms can move parallel or anti-parallel to each other. The latter case results in an order parameter smaller than 1. Hence, the mean order of 30 runs for $\eta=0$ is smaller than 1 (Fig.~\ref{fig:comparison_channel_PBCs}).

The standard deviation of $v_a$ for $\eta=0$ is relatively large (Fig.~\ref{fig:comparison_channel_PBCs}). Therefore, we performed parameter sweeps with 1000-4000 runs each, for systems with zero noise. These simulations showed that the mean of the order parameter is always in the range of $0.75 \pm 0.2 $ independent of system size, particle density or particle velocity.
Assuming that every possible state is equally probable, we derived the expectation value of the order parameter for zero noise as (see also Appendix \ref{Statistical solution for order in channel})

\begin{gather}
<v_a> = \frac{3N+2}{4N} \approx 0.75
\end{gather}
The obtained value of $<v_a>$ matches the mean value of $v_a$ from the simulations. This supports the observation that the system behaviour does not depend on the model parameters system size, particle velocity, and particle density.    

\subsection{Box} \label{Box}

As the second geometry, we consider a square box with reflective boundary conditions at all four sides. Hence, the particles are confined to a finite space. Long-lived velocity correlations measured by the polar order parameter $v_a$ of the base model cannot occur, because the ordering mechanism is constantly disrupted by particle-boundary collisions.  

\begin{figure}[ht!]
	\subfloat[\label{box:high_noise}] {%
		\includegraphics[ width= 0.48\linewidth ]{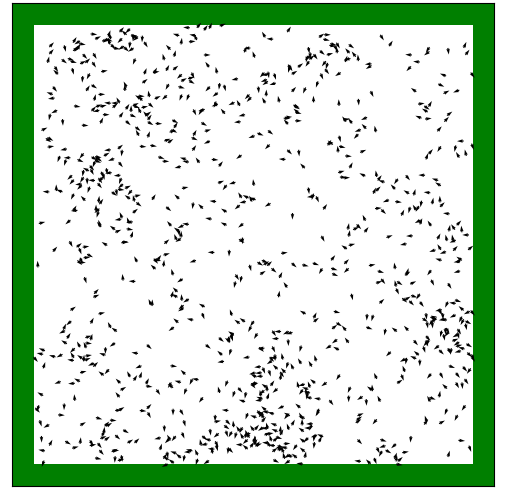}}
	\hspace{\fill}
	\subfloat[\label{box:low_noise}]{%
		\includegraphics[ width=0.48\linewidth ]{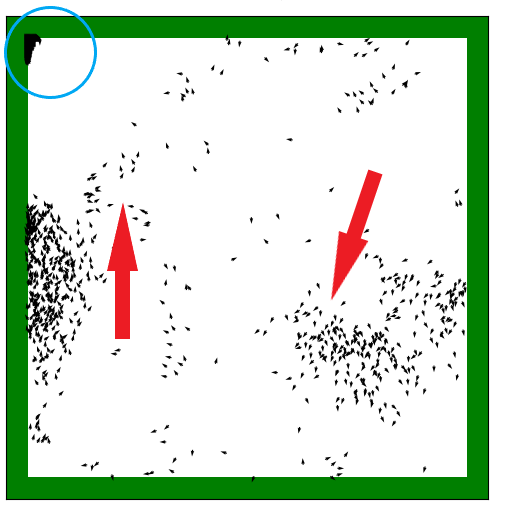}}
	\hspace{\fill}
	\caption{\label{fig:snapshot_box}Snapshots of the VM in a box geometry after 500 time steps, with $N = 1000, v_0 = 5, L = 500$, and $R = 18$ for noise strength $\eta = 3.2$ (a) and $\eta = 1.6$ (b). The red arrows indicate the movement direction of the two main swarms on the left and right boundary. Overall this results in a noisy clockwise rotational motion. The blue circle marks a trapped swarm in the top left edge. The reflective boundaries in the x- and y-direction are depicted as green walls. The black arrow heads indicate the current movement direction of the individual particles. }
\end{figure}

We performed simulations for different noise strength $\eta$. For high levels of noise, we did not observe particle swarms (Fig.~\ref{box:high_noise}). For low levels of noise, swarms form and move in arbitrary directions. 

The corners of the box disrupt the parallel alignment to the boundaries. If a swarm hits a corner, several things can happen. The most probable outcome is that a swarm aligns itself to the boundary that is perpendicular to its previous movement direction (see Fig \ref{fig:snapshot_croner_interactions}a-f). It is also possible that a swarm is trapped in the corner (Fig.~\ref{box:low_noise} and \ref{fig:snapshot_croner_interactions}g-l). Overall, this induces a noisy rotational movement along the system boundaries (Fig.~\ref{box:low_noise}). We performed a statistical analysis which showed that there is no preferred direction of the rotational motion. Swarms can align themselves clockwise or counter-clockwise with equal probability. Mixed rotational states are only intermediates that are resolved over time. 

Swarm trapping in corners is particular to simulations with low noise in the box geometry with reflective boundaries. If a swarm is directly approaching a corner, the first incoming particles can only get reflected to places within the interaction range of the other particles of the swarm due to the confined space. These reflected particles completely reverse their orientation in the next time step (Fig.~\ref{fig:snapshot_croner_interactions}) due to the alignment with the other still incoming particles of the swarm . Hence, the whole swarm does not change its direction over time, because only a fraction of the particles is reflected in each time step and their direction is quickly reversed. In addition, if the velocity of a particle is not sufficient to reach the back-end of the incoming swarm, the reflected particles accumulate in the front part of the swarm. This causes the volume of the swarm to decrease, but the density to increase. In the case of a finite volume of each particle (as in this simulation) this "concentration" of the swarm stops, if the closest packing determined by the lattice spacing is reached and no further movement is possible. For particles without spatial extension, another outcome is possible. The swarms can be compressed so much that the distance to the boundary is smaller than $v_0*\Delta t$ for all particles. Hence, in the following time step all particles of a swarm can collide simultaneously with the boundary and completely reverse their orientation and leave the corner. 

A trapped swarm has the effect of a sink on free particles or swarms, such that they are trapped as well, if they move into the interaction area of a trapped swarm.
 
\begin{figure}[htbp] 
	\centering
	\includegraphics[width=\linewidth]{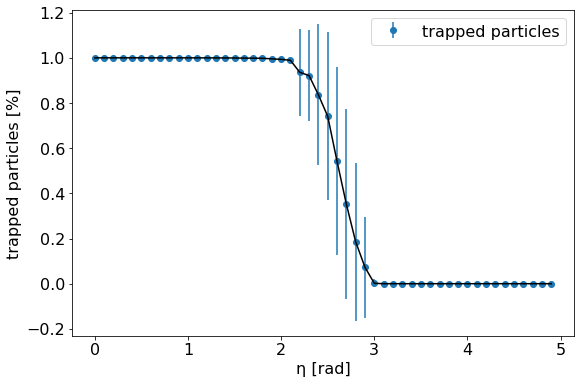}
	\caption{Percentage of trapped particles versus noise strength $\eta$ for a system with a box geometry after 15000 time steps, with $N = 2000, \rho = 2, v_0 = 5$ and $R =18$. The value of $\eta$ was increased from $0.0$ to $0.5$ in steps of $0.1$. Data points are the mean values with standard deviations of 30 runs each. The black lines connect the data points for a better visual perception of the transitions.}
	\label{fig:trapped_pa}
\end{figure}

Our observations suggest a dependence of the trapping behaviour on the noise strength $\eta$. We introduce the percentage of trapped particles as a metric to quantify the system. An analysis of the percentage of trapped particles after 15000 time steps for increasing $\eta$ shows that for $\eta < 2.1$ all particles get trapped eventually (Fig.~\ref{fig:trapped_pa}). For noise strength between $\eta = 2.2$ and $\eta = 3.0$, the percentage of trapped particles drops very fast to zero. For higher noise strengths, no particles get trapped at all.

\begin{figure}[htbp] 
	\centering
	\includegraphics[width=\linewidth]{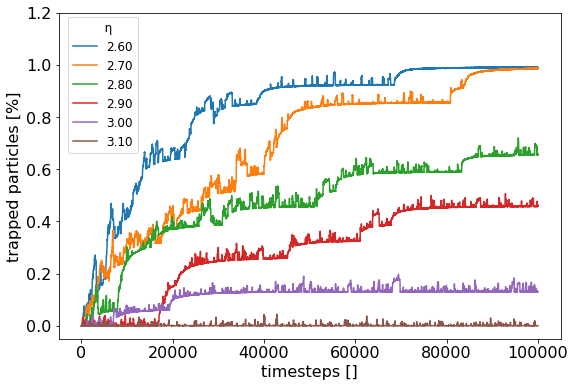}
	\caption{Time scaling of the percentage of trapped particles for different noise strengths $\eta$ for a system with a box geometry with $N = 2000, \rho = 2, v_0 = 5$ and $R =18$. Data points are the mean values of 15 runs each. For a better visual perception, the standard deviation has been omitted. Please refer to Fig. \ref{fig:trapped_pa_overtime_error} for full information.}
	\label{fig:trapped_pa_overtime}
\end{figure}

 The timescales for particle trapping are also worth investigating. In the VM with periodic boundary conditions, systems quickly reach a state where the order parameter $v_a$ and therefore the qualitative behaviour does not change anymore \cite{Vicsek}. The time for this "thermalisation" process depends on the relation of system size $L$, particle velocity $v_0$ and particle density $\rho$. For the parameter values used in this work, the final state is reached after at most 1000 time steps. In the case of the box geometry, the percentage of trapped particles does not reach a final state after a comparable time span. Therefore, we performed simulations for 100 000 time steps and calculated the percentage of trapped particles for different noise strength (Fig.~\ref{fig:trapped_pa_overtime} and Fig.~\ref{fig:trapped_pa_overtime_error}).    

For noise strength $\eta \leq 2.7$, all particles are trapped at the end of the simulation. For sufficiently high noise of $\eta = 3.1$, no particles are trapped. Hence, compared to the shorter simulations ( Fig.~\ref{fig:trapped_pa}), the noise interval for which only a part of particles get trapped, decreases to $\eta = 2.8$ - $\eta = 3.0$. Based on the shape of the graph for this noise regime, we suspect that the trapping rate has not reached a plateau, yet. Hence, we expect that for very long time spans, there is a discontinuous transition from no particle trapped to all particles trapped at $\eta \approx 3.1$.

\subsection{Disk}
\begin{figure}
	    \subfloat[\label{circle_low}]{%
		\includegraphics[ width=0.49\linewidth]{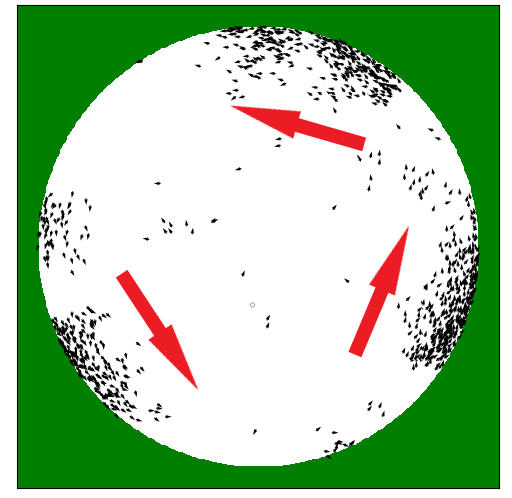}}
	\hspace{\fill}
	    \subfloat[\label{circle_high}] {%
		\includegraphics[ width=0.49\linewidth]{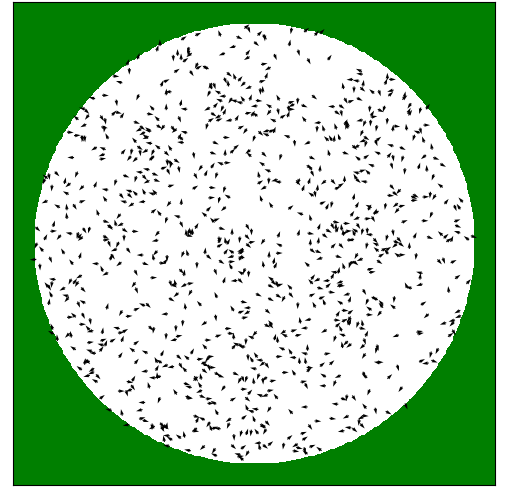}}
	\hspace{\fill}
\caption{\label{fig:snapshot_circle}Snapshots of the VM in a disk geometry after 500 time steps with $N = 1000, v_0 = 5, L = 500, R = 18$. (a) For a low noise strength  $\eta$ = 0.6, the system is in a rotational state. The red arrows indicate the movement direction of the three big swarms, which overall result in a counterclockwise rotation. (b) For a high noise strength $\eta$ = 3.2, the system does not exhibit swarms. The reflective boundaries are depicted as green walls. The arrow heads indicate the current movement direction of the individual particles. }
\end{figure}

We further analysed the behaviour of the VM in a disk geometry with reflective boundaries. Simulations for different levels of noise strength $\eta$ reveal a similar behaviour as in the other geometries. For low noise, the particles form swarms and with increasing noise the particle movement becomes uncorrelated (Fig.~\ref{fig:snapshot_circle}). 

\begin{figure}
\begin{center}
\includegraphics[width=0.75\linewidth]{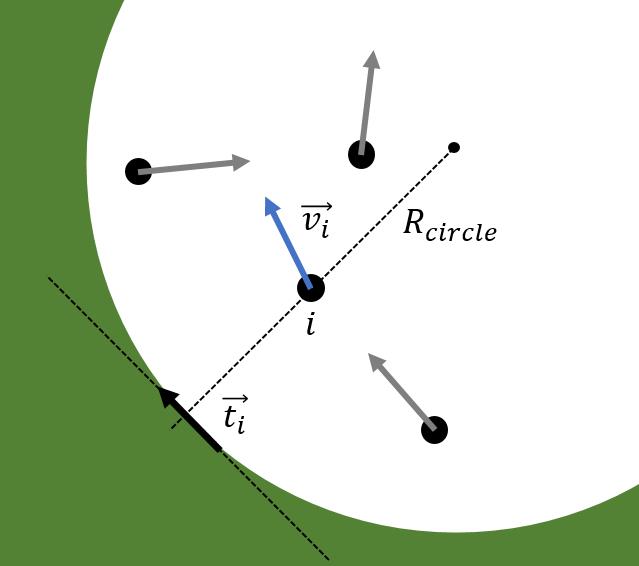}
\caption[]{Illustration of the order parameter $v_c$ for circular boundaries. For each particle, the scalar product of the velocity $\vec{v_i}$ and the tangential vector $\vec{t}_i$ of its projected position at the boundary, is calculated. The reflective boundaries are depicted as green walls.}   
\label{fig:circle_cor}
\end{center}
\end{figure}

As in the box geometry, the swarms align themselves to the system boundaries and eventually a stable rotational state is formed. Due to the steady curvature of the circular boundaries, the rotation is more regular than in the box geometry. As there is no long term alignment in one direction but alignment to the curved boundaries, the order parameter $v_a$ is not a suitable metric for this system. Therefore, motivated by \cite{Hiraoka_2017}, we constructed a new order parameter $v_c$ with     
\begin{equation}
	v_{c} = \frac{1}{Nv_0}\Bigg|\sum_{i=1}^{N}\frac{\vec{v}_i*\vec{t}_i}{|t_i|}\Bigg|,
	\label{eq:order_parameter_circle}
\end{equation}
where $\vec{t}_i$ is the tangent vector of the nearest boundary to the particle (Fig.~\ref{fig:circle_cor}). In comparison to the order parameter $v_a$ of the base model, the velocity is replaced by the scalar products of the particle velocity with the tangent vector $t_i$ of the nearest surface.

If $v_c = 1$, all particles are aligned parallel to their nearest boundary (Fig.~\ref{circle_low}), which is a perfect rotational state.  For large noise strength $\eta$, particle swarms and rotations are not observed and the particles move uncorrelated (Fig.~\ref{circle_high}). In this case, both $v_a$ and $v_c$ are equal to zero and equivalent descriptions of the state of the system.       

\begin{figure*}
	    \subfloat[\label{circle_order_1}]{%
	    \includegraphics[ width=0.49\textwidth]{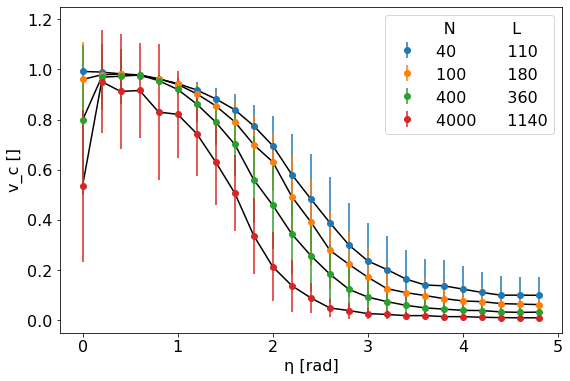}}
	\hspace{\fill}
	    \subfloat[\label{base_order_1}]{%
	    \includegraphics[ width=0.49\textwidth]{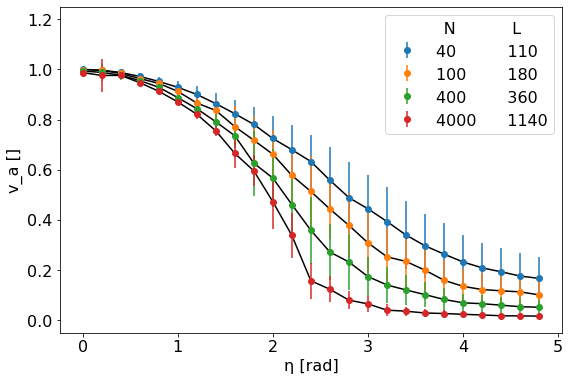}}

	\hspace{\fill}
	    \subfloat[\label{circle_denisty}] {%
	    \includegraphics[ width=0.49\textwidth]{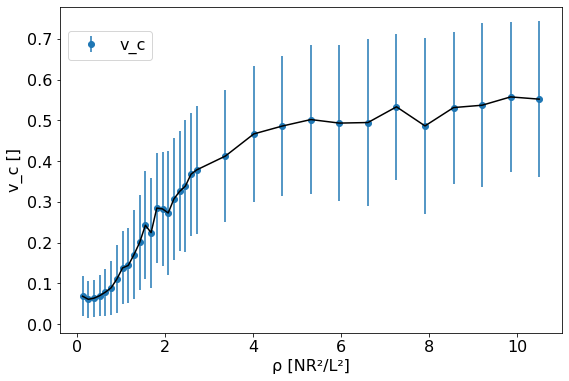}}
	\hspace{\fill}
	    \subfloat[\label{base_denisty}] {%
		\includegraphics[ width=0.49\textwidth]{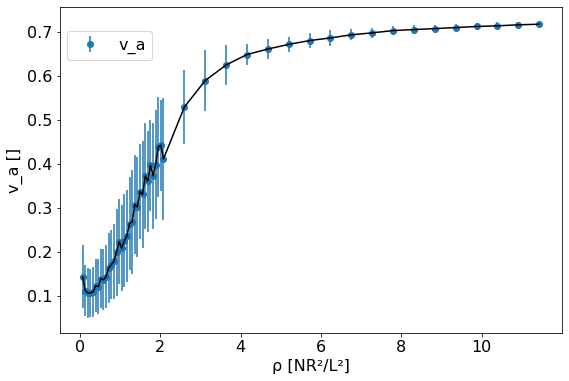}}
	\hspace{\fill}
	\caption{ Noise dependence with increasing system size $L$ and particle number $N$ for (a) the introduced circle order parameter $v_c$ in  a circle geometry with reflective boundary conditions and b) the order parameter $v_a$ in a similar sized box with periodic boundary conditions. The other model parameters are similar in all shown systems ($\rho = 1.5$, $v_0 = 5$ and $R =18$).\\
	Density dependency of (c) the circle order parameter $v_c$ in a circle geometry with reflective boundary conditions and of (d) the order parameter $v_a$ in a square with periodic boundary conditions. We chose systems with equal dimensions ($L = 400, v_0 = 5$ and $R =18$) and a fixed noise $\eta$ = 2.6. Data points are the mean values with standard deviation of 30 runs each.}
	\label{fig:circle_size_scaling_comparison}
\end{figure*}

The order parameter $v_c$ for circular boundaries shows a steady transition from high order at low noise to low order at high noise (Fig.~\ref{circle_order_1}) in a disk geometry. This is a very similar behaviour to the order parameter $v_a$ (Fig.~\ref{base_order_1}) in a square with periodic boundary conditions. 
In large systems and for zero noise, the mean of $v_c$ is smaller than for low noise and has a larger standard deviation. In this case, it can happen that the system freezes into a state, in which the swarms are trapped at the curved surface, in a very similar way as to the corners of a box geometry (see Fig.~\ref{fig:snapshot_stuck_circle}). In these frozen states, it is possible that states of two (or more) oppositely oriented swarms (clockwise and counterclockwise) survive long-term. If so, the circle correlation function $v_c$ assumes values lower than 1. This explains the smaller mean value of $v_c$ and its high standard deviation for zero noise.  This trapping phenomenon appears to be an effect of the combination of an on-lattice model with zero noise. We expect that for a perfectly continuous circular curvature and zero particle volume this effect should not occur.

A general trend for different system sizes is that $v_c$ for the disk geometry decreases faster to zero than $v_a$ for the base model with increasing noise $\eta$ (Fig.~\ref{circle_order_1}). Fixing the noise to a medium strength of $\eta=2.6$ and varying the density confirms that $v_c(\rho)<v_a(\rho)$ (Fig.~\ref{circle_denisty} and \ref{base_denisty}).

\section{Discussion}
We introduced an on-lattice version of the VM. The inherently included finite volume of particles and the easy expandability make this model attractive for an application to a wide range of real life systems.
We employed simulations with reflective boundary conditions for different geometries and studied the effect of these alterations on the behaviour of the model. For a sufficiently fine lattice, the on-lattice implementation does not change the ordering behaviour measured by the correlation function $v_a$. Even though, we are certain that the used implementation does not change the examined model dynamics, we cannot fully rule out that other metrics like fluctuation correlations or number fluctuations \cite{Number_fluctuations} could reveal subtle difference between different implementations. Future studies could take a closer look at that.
The introduction of reflective boundaries to the VM drastically changes its macroscopic behaviour. Whereas the direction of the final macroscopic state in the base model is randomly chosen, the combination of reflective and periodic boundary conditions in a channel geometry forces the direction of the final state to be parallel to the reflective boundaries. Upon swarm collision with a boundary, the swarm particles velocity components perpendicular to the boundary are cancelled and the particles align their velocity parallel to the boundary. We infer that the circular interaction area of a Vicsek particle is primarily responsible for this behaviour. If the interaction area of a particle would be limited to particles in front, this effect would not occur. In some active matter systems (eg. fish, birds,...) where the eyes, and therefore the alignment, are primarily directed upfront, such a change to the model appears to be justified \cite{Ducks_swimming} and could be worth investigating.
Different to an analysis of the continuous VM in a similar geometry, we did not observe an oscillating shear flow state \cite{ARMBRUSTER201758}. The discrete VM relies on an instantaneous direction alignment, while in the continuous version the alignment is smooth and gradual \cite{Vicsek_continous}. We suspect that the instantaneous alignment suppresses the oscillating states.\\
In a box with reflective boundaries, we found that aligned particles can get trapped in the corners. We concluded that this effect is a consequence of a finite particle volume. In the continuous version of the VM, the particles do not possess a finite volume.  Off-lattice simulations of the VM variant in a similar box geometry have not exhibited particle trapping \cite{ARMBRUSTER201758}. \\
The transition between states where all particles are free and all particles are trapped appears to be discontinuous at $\eta \approx 3.1$. This behaviour is reminiscent of the phase transition from order to disorder in the basic VM and preliminary parameter sweeps indicate a similar dependence of the two transitions on the parameters $\rho,v_r,L$. In addition, the transition from the free to the trapped behaviour shows surprising similarities with a first order phase transition from liquid to solid in thermal equilibrium systems, where $\eta$ plays the role of the temperature, and the percentage of trapped particles the role of an order parameter.  However, a more detailed systematical study of the transition properties is required to fully answer these questions.  \\
Our analysis of the VM in a disk geometry shows a rotational state for low noise strength and uncorrelated movement for high noise strength. These results are in agreement with previous off-lattice simulations of the VM \cite{CZIROK200017}. For the off-lattice implementation, an additional local repulsion has been introduced to avoid particle overlap. This effect comes naturally with our on-lattice implementation. Our results are also consistent with the behaviour described for the continuous VM in a disk geometry \cite{ARMBRUSTER201758}. To quantify the rotational state, we introduced an order parameter $v_c$. We find that the order described by $v_c$ decreases with increasing noise level and increases with increasing particle density. Quantification of a system with repulsive particles in a disk geometry with a comparable order parameter have shown similar results \cite{Hiraoka_2017}. Hence, overall our findings suggest that an ordered rotational state is universal in active matter systems in a disk geometry.\\
Comparison of the order parameter $v_c$ for a disk geometry and $v_a$ for a square with PBCs show a similar dependence on the noise strength $\eta$ and the particle density $\rho$. For equal densities, noise or system sizes, $v_c$ for a disk is smaller than $v_a$ for PBCs. This is indeed reasonable. If we assume a perfectly ordered state in both systems and zero noise. Then in the system with periodic boundary conditions, all particles move in the same direction. There is no additional alignment between the particles needed to preserve that state. In a disk geometry, a perfectly ordered rotational state means that all particles are aligned to the tangential of the closest point on the boundary. As described for the channel geometry (section~ \ref{Channel}), the boundary alignment is due to a combination of reflective boundary conditions and the Vicsek alignment (eq.~\ref{eq:direction}). Hence, recurrent alignment between the particles and the boundary is required to preserve the ordered rotational state. Therefore, all parameter variations that weaken the alignment mechanism (eg. increased noise, decreased density) also weaken the rotational order in the disk to a larger extent than the order in a system with periodic boundary conditions. This difference between order in the two systems is well captured by the two parameters $v_a$ and $v_c$. 

In summary, we performed simulations for the VM on a lattice for a channel, a box and a disk. The on-lattice implementation did not effect the behaviour of the particle in the Model, while the reflective boundary conditions yield a velocity alignment of the particles parallel to boundaries for sufficiently low levels of noise. In the case of the box geometry, this can lead to particle trapping in the corners. 
The three geometries are each relevant for application to different biological systems. Movement of pedestrians or insect groups such as ants is often restricted by fences or walls that limit a channel \cite{ants_&_humans}. Collective behaviour of unicellular organisms is typically studied in quasi two-dimensional disk-like liquid droplets \cite{Krueger_QuantSoMo_2021} or in microfluidic devices with channels as well as chambers of different shapes including disks and boxes \cite{Liu_microfluidics_2011}. Our work provides a further step towards the application of the VM to study such systems.

\begin{acknowledgments}
We thank Wolfgang Kinzel and Holger Stark for fruitful discussions.
\end{acknowledgments}

\section*{Author contribution statement}
AK performed the computer simulations und data analyis. SCF supervised the research. AK and SCF wrote the manuscript.

\section*{Data availability}
The model implementation in Python together with the data sets generated and analysed during the study are available from the corresponding author on reasonable request.

\pagebreak
\bibliography{apssamp}

\newpage
\appendix
\beginsupplement
\section{Comparison on- and off-lattice implementation} \label{Comparison_on_off}

\begin{figure}[ht!]
	\subfloat[\label{circle_grid}]{%
		\includegraphics[ width=0.35\linewidth ]{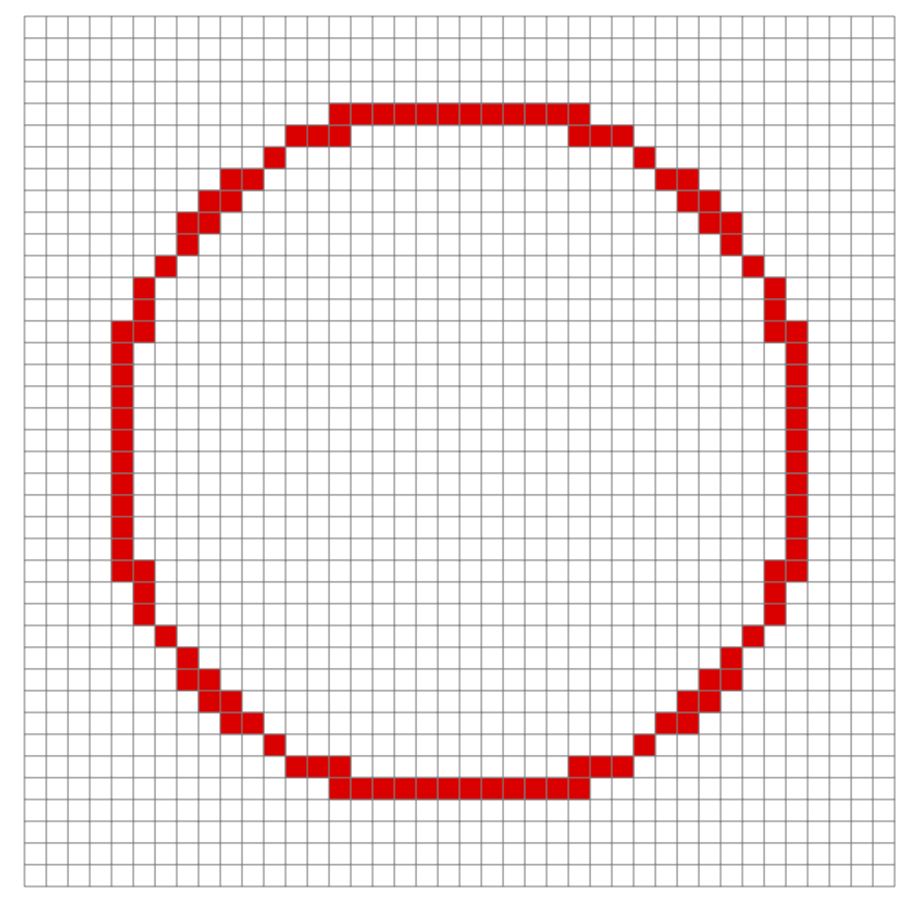}}
		\hspace{\fill}
	\subfloat[\label{orderpara_discretisation}] {%
	    \includegraphics[ width= 0.99\linewidth ]{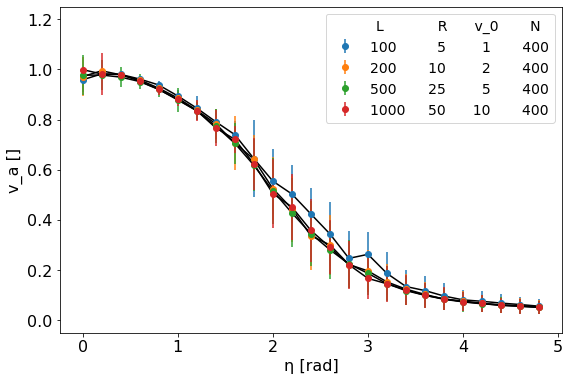}}
    \caption{\label{fig:comparision_lattice_2} a) Approximation of the circular interaction area on a grid for an interaction radius $R=15$. 
		\\ b) Order parameter $v_a$ versus the noise strength $\eta$ for systems with identical densities $\rho = \frac{NR^2}{L^2}$ and increasingly fine lattices. Data points are the mean values with standard deviations of 50 runs each. The black lines connect the data points for a better visual perception of the relations shown.\\
}
\end{figure}

The discretization influences two aspects of the model. Firstly, all particle positions (starting positions and those subsequently calculated by eq.~\ref{eq:next_step}) are rounded to the next discrete grid point. The relative error of  rounding decreases with increasing $v_0$. Secondly, approximation of the interaction area with discrete grid points causes a deviation from a perfectly circular region (eq.~\ref{eq:direction}) (see Fig.~\ref{circle_grid}). The relative error of this decreases with increasing $R$.

Performing simulations for systems with increasingly fine lattices and otherwise identical parameter values shows that only in the case of $v_0 =1$ (one unit cell) the behaviour is different (Fig.~\ref{orderpara_discretisation}). We expect that this deviation is caused by rounding artefacts. Investigations of the order parameter relative to noise strength as well as density shows identical behaviour as in the original publication of Vicsek et. al \cite{Vicsek} (Fig.~\ref{base_order_1} and \ref{base_denisty}). Therefore, we conclude that for a sufficiently fine lattice ($v_0 \geq 2$), the model behaviour is independent of the type of implementation.

\section{Statistical derivation for order in the channel geometry} \label{Statistical solution for order in channel}
This section contains a statistical derivation of the expectation value of the order parameter $v_a$ in a channel geometry for zero noise. As mentioned in section \ref{Channel}, the system can get "trapped" into a state where all particles are perfectly aligned to the walls and the particles at the top and the bottom wall do not interact with each other. In this case, four possible "macro" configurations can exist:
\\
1. All particles move parallel to the walls to the right (we call this parallel)\\
2. All particles move parallel to the walls to the left (we call this anti-parallel)\\
3. Particles on top move parallel and particles on the bottom move anti-parallel \\
4. Particles on top move anti-parallel and particles on the bottom move parallel \\
\\
Hence, the particles have two possible velocities $(v_0,0)$ or $(-v_0,0)$. Therefore, equation \ref{eq:order_parameter} can be simplified to:
\begin{equation}
v_a = \frac{1}{N}\Bigg|\sum_{i=1}^{N}d_i\Bigg|,
\label{eq:order_parameter_mod}
\end{equation}
where $d_i=1$ for a parallel moving particle and $d_i=-1$ for an anti-parallel moving particle. For each of the macro configurations, $N$ micro states are possible ($N$ particles at the top $\vert$ 0 particles at the bottom, ..., $N-j$ particles at the top $\vert$ $j$ particles at the bottom, ..., 0 particles at the top $\vert$ $N$ particles at the bottom). In the first two macro configurations, every micro state gives rise to the same order parameter ($v_a= 1$). In the third and fourth macro configuration, every micro state gives a different order parameter. If we assume that all of these $4N$ micro states are equally probable, the expectation value of the order parameter can be calculated by summation over all micro states:       
\begin{eqnarray}
<v_a> =&& \frac{1}{4N}(N+N +\frac{1}{N}\sum_{j=0}^{N}|-j+(N-j)|\nonumber \\
&&+\frac{1}{N}\sum_{j=0}^{N}|j-(N-j)|)
\end{eqnarray}
The normalisation originates from the $4N$ possible micro states. Each of the four summands is the sum over the order parameter of all micro states of one macro configuration. For the first two macro configurations, each micro state has maximum order ($v_a= 1$). Therefore, the sum can be simplified to $N$. This is not the case for the "mixed" configurations 3 and 4. Here, $j$ is the number of particles at the bottom, which is positive for parallel moving particles and negative for anti-parallel moving particles. This can be simplified to: 
\begin{gather}
<v_a> = \frac{1}{4N^2}(2N^2 +2\sum_{j=0}^{N}|-j+(N-j)|) \\
= \frac{1}{4N^2}(2N^2 +2\sum_{j=0}^{N}|-2j+N)|) \\
= \frac{1}{4N^2}(2N^2 +2\sum_{j=0}^{N/2}|2j|) \\
= \frac{1}{4N^2}\Bigg(2N^2 +2\Bigg(\frac{N}{2}\Bigg(\frac{N}{2}+1\Bigg)\Bigg)\\
= \frac{3N+2}{4N} \approx 0.75
\end{gather}

For large values of $N$ the expectation value of $v_a$ is approximately $0.75$ which corresponds well with the simulations.

\begin{figure*}
	    \subfloat[\label{reflect02}]{%
		\includegraphics[ width=0.16\textwidth]{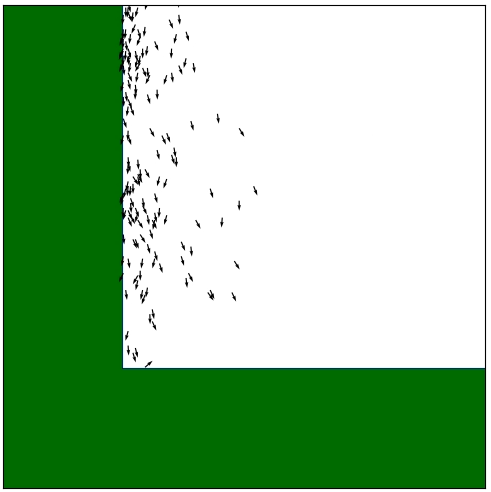}}
	\hspace{\fill}
	    \subfloat[\label{reflect03}]{%
		\includegraphics[ width=0.16\textwidth]{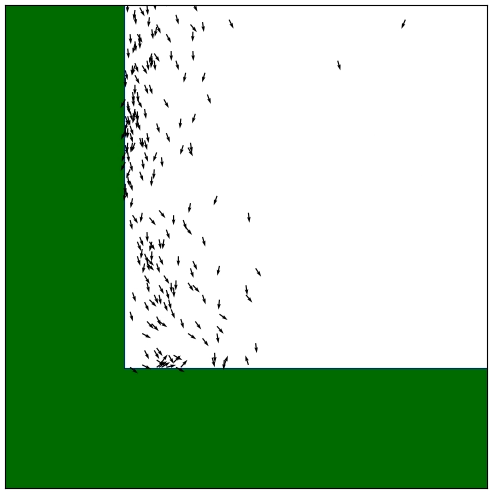}}
	\hspace{\fill}
		    \subfloat[\label{reflect04}]{%
		\includegraphics[ width=0.16\textwidth]{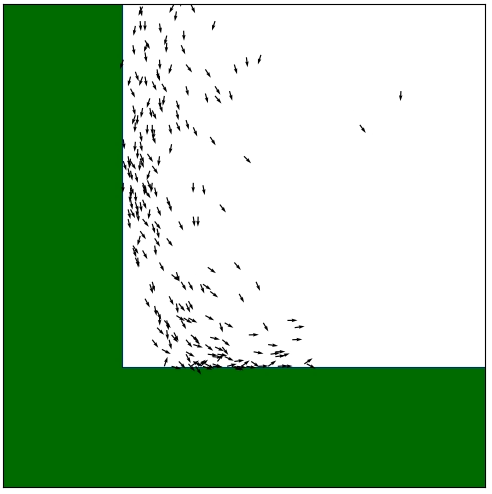}}
	\hspace{\fill}
		    \subfloat[\label{reflect05}]{%
		\includegraphics[ width=0.16\textwidth]{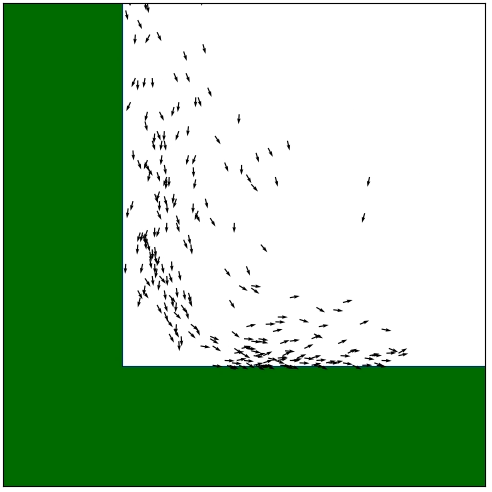}}
	\hspace{\fill}
		    \subfloat[\label{reflect06}]{%
		\includegraphics[ width=0.16\textwidth]{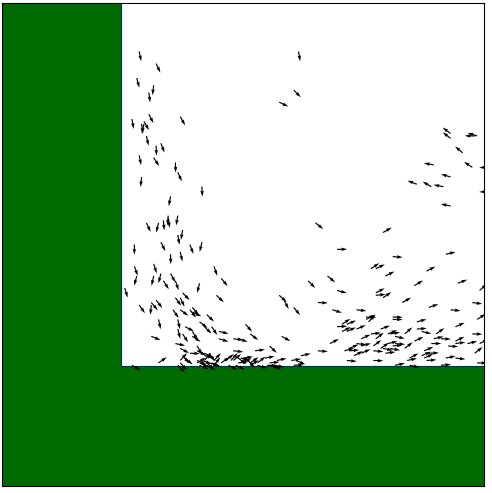}}
	\hspace{\fill}
		    \subfloat[\label{reflect07}]{%
		\includegraphics[ width=0.16\textwidth]{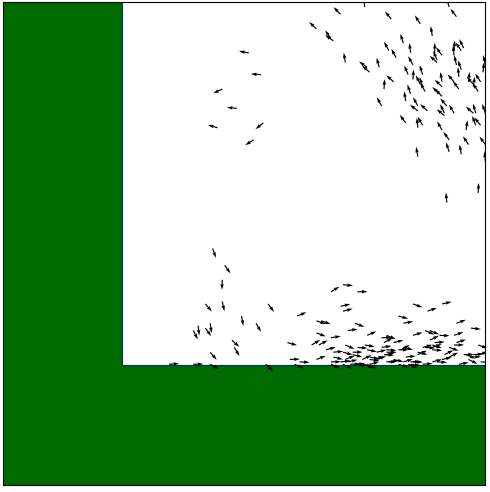}}
	\hspace{\fill}
		    \subfloat[\label{stuck02}]{%
		\includegraphics[ width=0.16\textwidth]{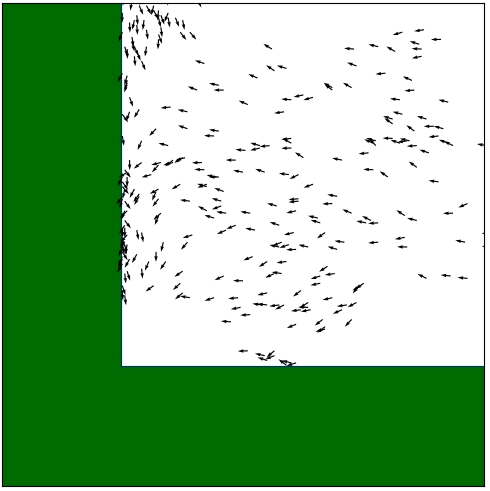}}
	\hspace{\fill}
	    \subfloat[\label{stuck03}]{%
		\includegraphics[ width=0.16\textwidth]{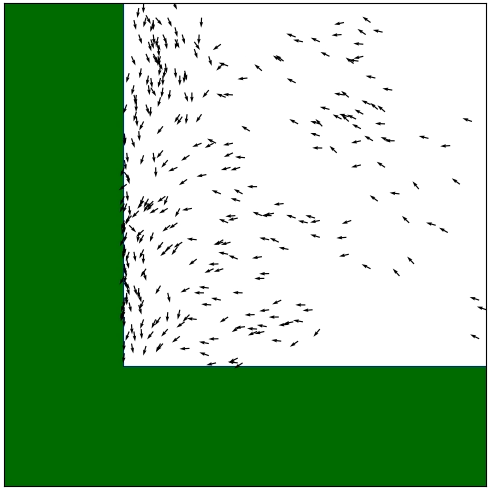}}
	\hspace{\fill}
		    \subfloat[\label{stuck04}]{%
		\includegraphics[ width=0.16\textwidth]{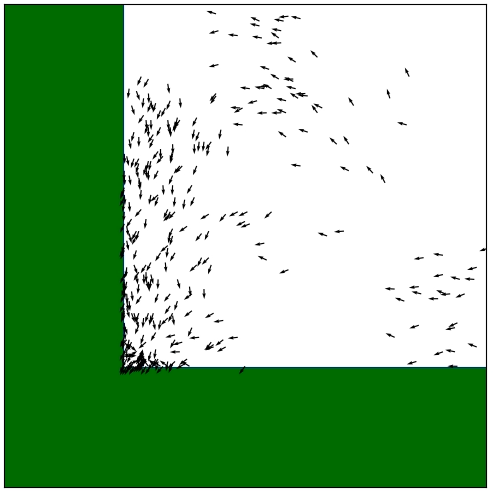}}
	\hspace{\fill}
		    \subfloat[\label{stuck05}]{%
		\includegraphics[ width=0.16\textwidth]{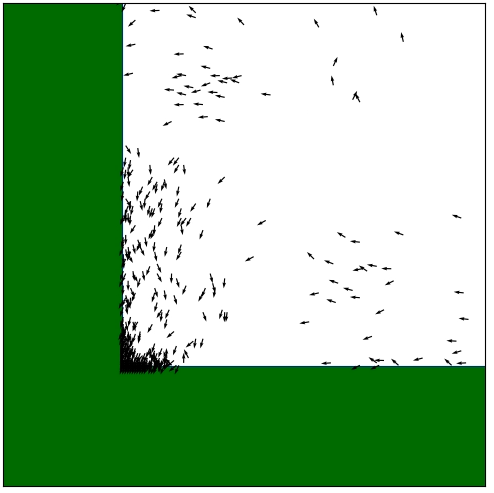}}
	\hspace{\fill}
		    \subfloat[\label{stuck06}]{%
		\includegraphics[ width=0.16\textwidth]{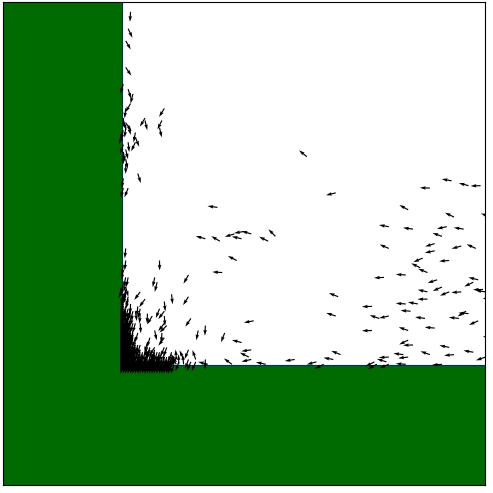}}
	\hspace{\fill}
		    \subfloat[\label{stuck07}]{%
		\includegraphics[ width=0.16\textwidth]{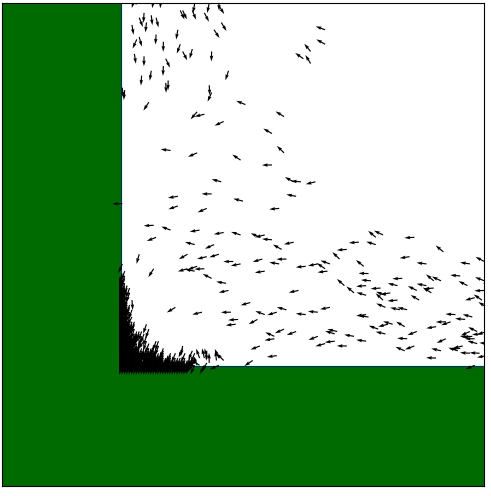}}
\caption{\label{fig:snapshot_croner_interactions} Consecutive snapshots every five time steps of the corner of a box geometry ($N = 4000, v_0 = 5, L = 1000, R = 18$). The arrow heads indicate the current movement direction of the individual particles. The reflective boundaries are depicted as green walls.  (a-f) Sequence 1: A swarm approaches the corner from the left edge and gets reflected alongside the bottom edge. 
	(g-l) Sequence 2: Swarms approach the corner from the middle and left edge and get trapped in the corner. }
\end{figure*}

\begin{figure}
    \centering
    \includegraphics[ width=1.0\linewidth]{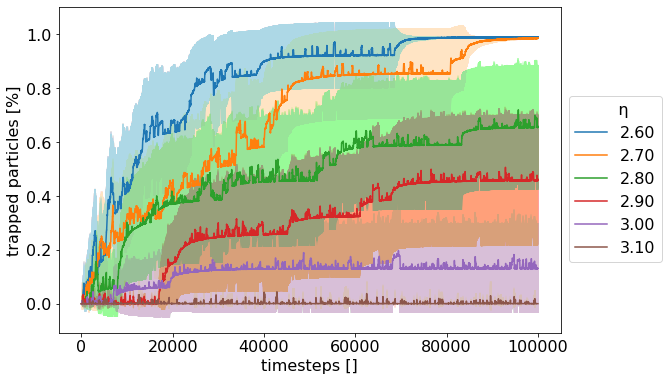}
	\caption{Time scaling of the percentage of trapped particles for different noise strengths $\eta$ for a system with a box geometry with $N = 2000, \rho = 2, v_0 = 5$ and $R =18$. Data points are the mean values of 15 runs each. The shaded areas indicate the standard deviation.}
	\label{fig:trapped_pa_overtime_error}
\end{figure}

\begin{figure*}
	    \subfloat[\label{stuck_circle_01}]{%
		\includegraphics[ width=0.16\textwidth]{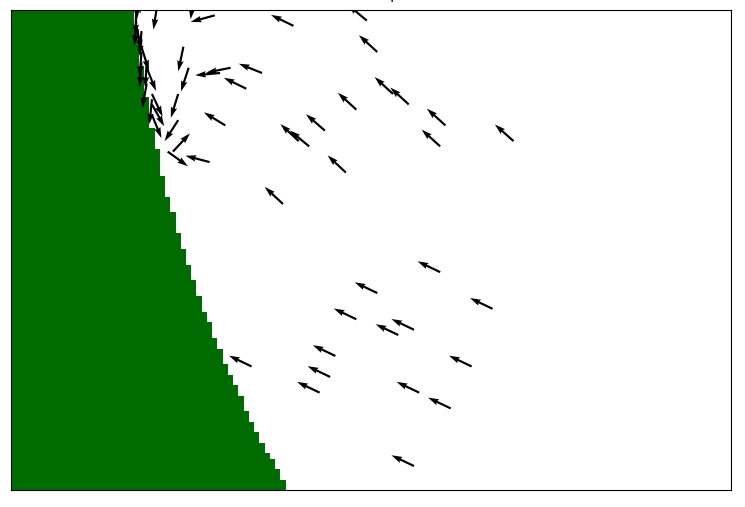}}
	\hspace{\fill}
	    \subfloat[\label{stuck_circle_02}]{%
		\includegraphics[ width=0.16\textwidth]{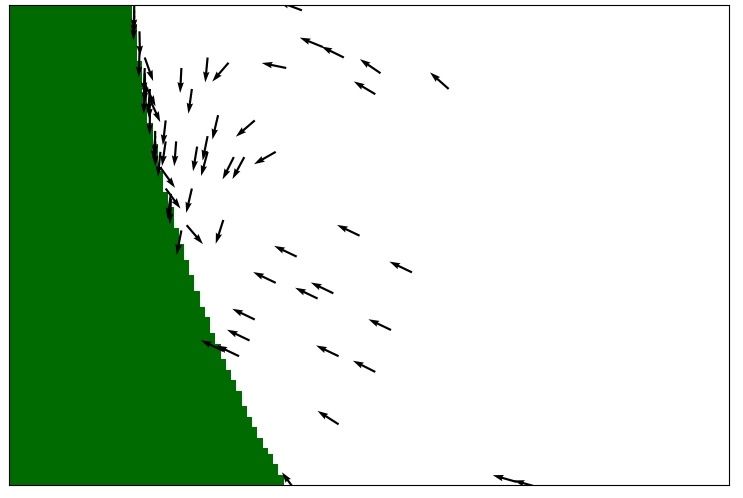}}
	\hspace{\fill}
		    \subfloat[\label{stuck_circle_03}]{%
		\includegraphics[ width=0.16\textwidth]{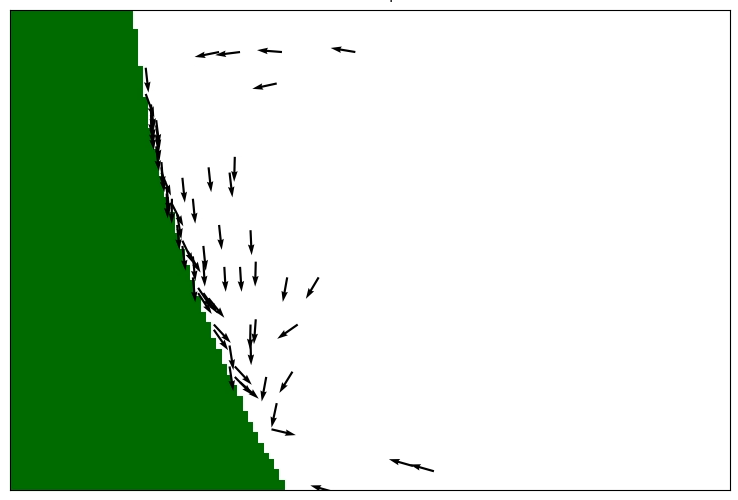}}
	\hspace{\fill}
		    \subfloat[\label{stuck_circle_04}]{%
		\includegraphics[ width=0.16\textwidth]{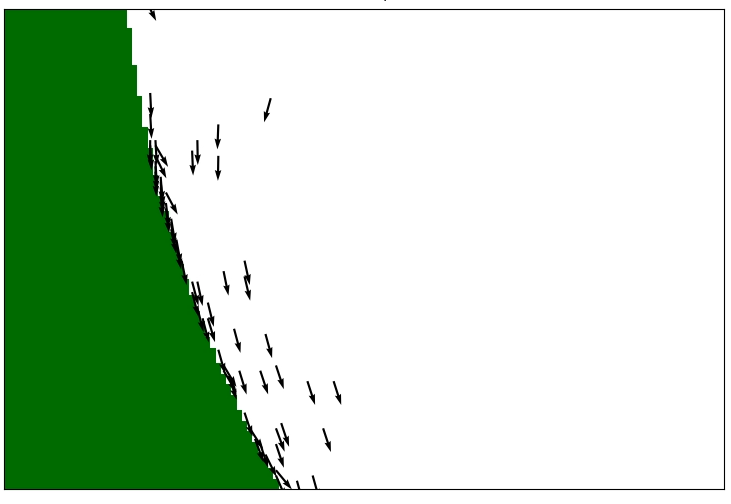}}
	\hspace{\fill}
		    \subfloat[\label{stuck_circle_05}]{%
		\includegraphics[ width=0.16\textwidth]{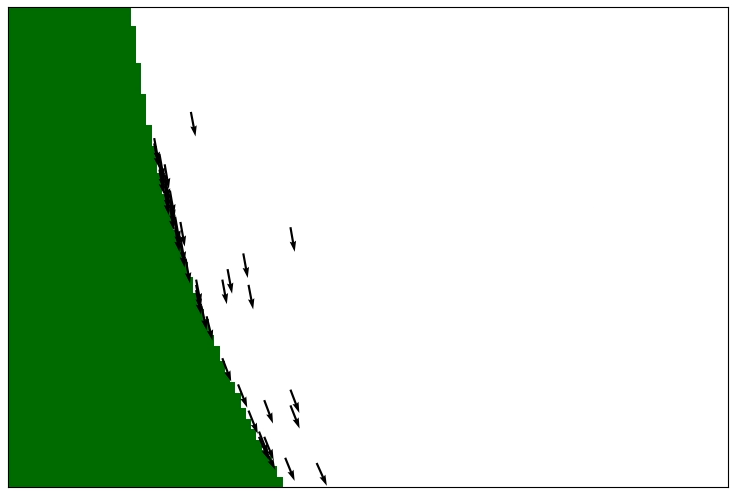}}
	\hspace{\fill}
		    \subfloat[\label{stuck_circle_06}]{%
		\includegraphics[ width=0.16\textwidth]{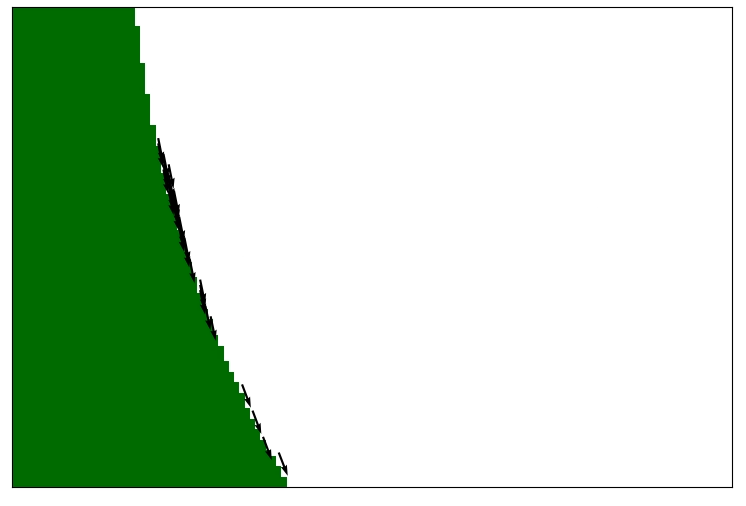}}

\caption{\label{fig:snapshot_stuck_circle} Consecutive snapshots every five time steps of a boundary section of a disk geometry ($N = 400, v_0 = 5, L = 400, R =18$) for zero noise ($\eta = 0$). The arrow heads indicate the current movement direction of the individual particles. The reflective boundaries are depicted as green walls.}
\end{figure*}

\end{document}